\documentclass[oldversion,rnote]{aa} 
\usepackage{graphicx}
\usepackage{aalongtable}
\usepackage{txfonts}
\usepackage{rotating}
\usepackage{lscape}
\newcommand{\gsim}{\;\lower.6ex\hbox{$\sim$}\kern-7.75pt\raise.65ex\hbox{$>$}\;}
\newcommand{\lsim}{\;\lower.6ex\hbox{$\sim$}\kern-7.75pt\raise.65ex\hbox{$<$}\;}

\begin{document}
\title{Aluminum abundances of multiple stellar generations in the globular
cluster NGC~1851\thanks{Based on observations collected at 
ESO telescopes under programme 188.B-3002}
 }

\author{
Eugenio Carretta\inst{1},
Valentina D'Orazi\inst{2},
Raffaele G. Gratton\inst{3},
\and
Sara Lucatello\inst{3}
}

\authorrunning{E. Carretta et al.}
\titlerunning{Aluminum abundances in NGC~1851}

\offprints{E. Carretta, eugenio.carretta@oabo.inaf.it}

\institute{
INAF-Osservatorio Astronomico di Bologna, Via Ranzani 1, I-40127
 Bologna, Italy
\and
Dept. of Physics and Astronomy, Macquarie University, Sydney, 
NSW, 2109 Australia 
\and
INAF-Osservatorio Astronomico di Padova, Vicolo dell'Osservatorio 5, I-35122
 Padova, Italy }

\date{}

\abstract{We study the distribution of aluminum abundances among red giants in
the peculiar globular cluster NGC~1851. Aluminum abundances were derived from 
the strong doublet Al {\sc i} 8772-8773~\AA\ measured on intermediate 
resolution FLAMES spectra of 50 cluster stars acquired under the Gaia-ESO
public survey. We coupled these abundances with
previously derived abundance of O, Na, Mg to fully characterize the interplay of
the NeNa and MgAl cycles of H-burning at high temperature in the early stellar
generation in NGC~1851. The stars in our sample show well defined correlations
between Al,Na and Si; Al is anticorrelated with O and Mg. The average value
of the [Al/Fe] ratio steadily  increases going from the first generation stars
to the second generation populations with intermediate and extremely modified
composition. We confirm on a larger database the results recently obtained by us
(Carretta et al. 2011a): the pattern of abundances of proton-capture elements
implies a moderate production of Al in NGC~1851. We find evidence of a
statistically significant positive correlation between Al and Ba abundances in
the more metal-rich component of red giants in NGC~1851.
  }

\maketitle

\section{Introduction}
The ubiquitous presence in galactic globular clusters (GCs) of multiple stellar
populations is amply assessed from the wealth of recent, high quality data (see
e.g. the reviews by Gratton et al. 2004, Martell 2011 and Gratton et al. 2012a). 
Spectroscopy, in particular (but in a few cases also the splitting of photometric 
sequences, see e.g. Piotto 2009 and Bragaglia 2010) does prove
beyond all doubt that the stellar population in GCs is composed of at least two
stellar generations, distinct in ages (with a slight age difference from a
few million years up to a few 10$^7$ years, depending on the nature of
polluters) and in particular in a chemical composition even hugely different 
between each other, although these two generations may not be often seen as 
discrete groups in the abundance planes and/or in the colour-magnitude diagram.

This chemical signature must be necessarily attributed to the action of the more
massive stars of an early, first generation since only this kind of stars could
be responsible for the present pattern of Na-O and Mg-Al anticorrelations
observed in present day unevolved cluster stars (Gratton et al. 2001), through
the nuclear processing of H burning at high temperature (Denisenkov \&
Denisenkova 1989, Langer et al. 1993).

Most evidence concerning multiple stellar populations, and the recently found
links with global cluster parameters (like total mass, Carretta et al. 2010, or
horizontal branch morphology, Gratton et al. 2010) mainly stems from the Na-O
anticorrelation, extensively studied thanks to modern facilities such as
FLAMES@VLT (e.g. Carretta et al. 2009a,b). This feature was soon realized to be
related to the intrinsic mechanism of formation of GC (Carretta 2006) and it is
so widespread among GCs that can be considered as the simplest working
definition of {\it bona fide} globular cluster (Carretta et al.
2010)\footnote{The Na-O anticorrelation is found in almost all GCs were Na,
O abundances are derived for a large number of stars, although some claims -
based on theoretical arguments - of single-generation clusters was made by e.g.
D'Antona \& Caloi (2008). As an example they suggest that NGC~6397 was composed
only by second generation stars. However, both Carretta et al. (2010, using
mainly Na) and Lind et al. (2011, with both Na and O) were able to show the
presence of two stellar generations also in this globular cluster.}.

However, since the study by Gratton et al. (2001), it was immediately clear the
importance of collecting the widest range of elements produced in proton-capture
reactions. In particular, the ``heaviest" light elements involved, such as Mg, 
Al and Si allow us to explore the hottest regime and cycles of the H burning.
This is crucial in order to shed light on the still uncertain nature of the
candidate polluters of the first stellar generation whose identification appears
still problematic and debated, with favourite candidates being fast rotating
massive stars (FRMA: Decressin et al. 2007) or intermediate mass asymptotic
giant branch (AGB) stars (Ventura et al. 2001)\footnote{In a few cases,
like  M~22 (Marino et al. 2009) and NGC~1851 (Carretta et al. 2011a), also a
contribution by type II SNe to the cluster pollution must be taken into
account.}.

To this aim we started to add Al abundances to the very large dataset with O,
Na, Mg, Si abundances already in hand, by re-observing large sample of stars of
our FLAMES survey of GCs. Proprietary data are being analyzed for a few key
objects (see the results on NGC~6752, Carretta et al. 2012). In the present note
we exploit the observations made in NGC~1851 by the Gaia-ESO
spectroscopic public survey, just started on FLAMES@VLT.

\section{Observations} 
We retrieved from the ESO archive two exposures, of 600 sec each made with
FLAMES mounted at VLT-UT2 and the high resolution grating HR21. 
The observations were carried out on UT 17 February 2012, with airmass
z=1.115 and 1.155, respectively. The resolution of HR21 is 17,300 and the
spectral range is from 8484~\AA\ to 9001~\AA, including the 
Al~{\sc i} doublet at 8772-73~\AA, which is the feature that we use.

Data reduction was performed through the ESO FLAMES-Giraffe pipeline (version
2.8.9, http://www.eso.org/sci/software/pipelines//giraffe/giraf-pipe-recipes.html),
which provides bias-corrected, flat-fielded,
1D-extracted and wavelength-calibrated spectra. Sky subtraction, combination of
the two single exposures for each star, and rest-frame traslation were then
carried out within IRAF\footnote{IRAF is the Image Reduction and Analysis
Facility, a general purpose software system for the reduction and analysis of
astronomical data. IRAF is written and supported by the IRAF programming group
at the National Optical Astronomy Observatories (NOAO) in Tucson, Arizona. NOAO
is operated by the Association of Universities for Research in Astronomy (AURA),
Inc. under cooperative agreement with the National Science Foundation.}.

\begin{figure}
\centering
\includegraphics[scale=0.40]{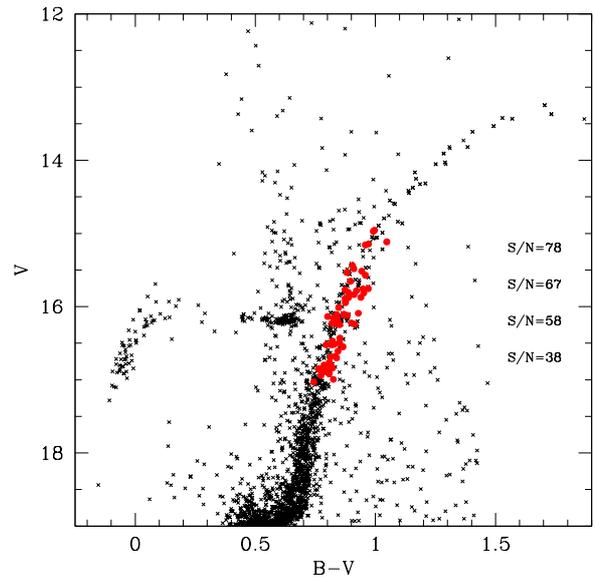}
\caption{$V,B-V$ colour-magnitude diagram (CMD) of NGC~1851 (grey crosses).
Superimposed as large red filled circles are the giants with Al abundances
measured in the present study. The average S/N ratio of spectra in bin of half a
magnitude is reported on the right of the CMD.}
\label{f:cmdsn}
\end{figure}

A total of 84 stars were observed in the two exposures, however we restricted
our attention to the sub-sample of 63 stars in common with the database by
Carretta et al. (2011a). For these stars we already have a full abundance
analysis, with homogeneously determined atmospheric parameters, and abundances
of light elements O, Na, Mg, Si. The spectral range of HR21 contains no useful
Fe lines, hence we restricted the present analysis to this subsample. 
It was not possible to measure reliable Al abundances for 13 stars that are too
hot (effective temperature higher than about 4900 K) or with low-quality
spectra. Hence the final sample with a complete set of light-element abundances
is represented by 50 red giant branch (RGB) stars, in a range of two magnitudes
(from  $V\sim 17$ to $V\sim 15$) approximatively centred at the luminosity of
the bump on the RGB ($V=16.16$, see Fig.~\ref{f:cmdsn}). Relevant information on
this final sample are listed in Table~\ref{t:tab1}.

The cluster was observed as calibrator for the survey, so the desired S/N was
set to the average one for survey stars, and not optimized to get a S/N ratio
good also for the much weaker Al lines.
The S/N ratios at 8800~\AA, as measured on the combined spectra, range from 
12 to 147 for the used sample, with a median value of 57. The average values of
the S/N in 0.5 mag bins are shown in Fig.~\ref{f:cmdsn}, whereas individual
values for each star are listed in Table~\ref{t:tab1}.

\begin{figure}
\centering
\includegraphics[bb=135 149 440 700,scale=0.80]{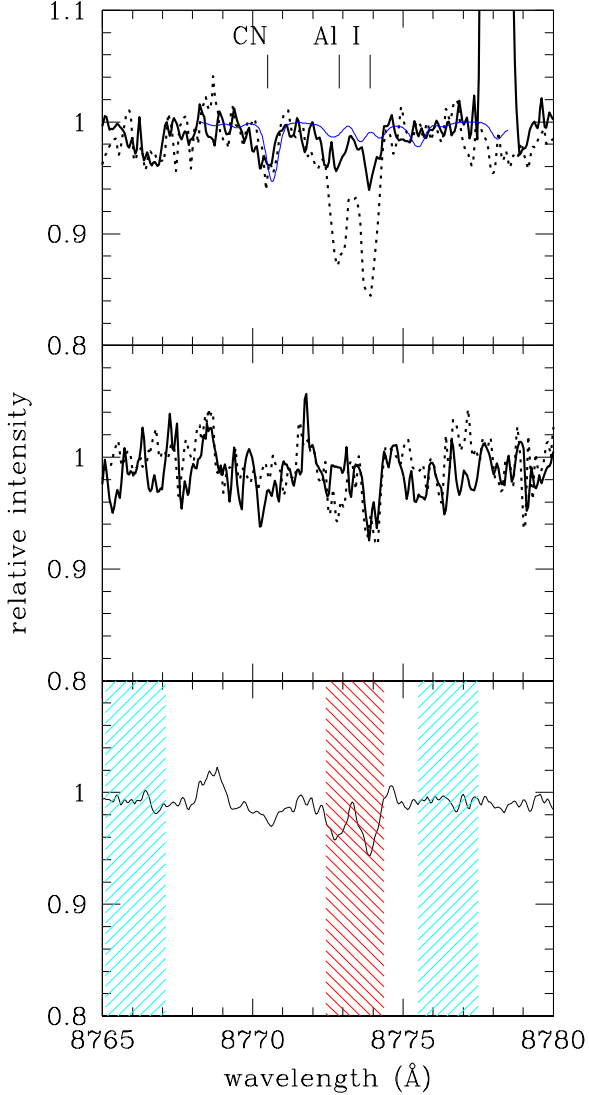}
\caption{Upper panel: comparison of the observed spectra for star 44939 (solid
line, [Al/Fe]=-0.19 dex) and star 35999 (dotted line, [Al/Fe]=+1.06 dex). 
The thin blue solid line is the synthetic spectrum with no Al ([Al/Fe]=-9.99)
computed with the atmospheric parameters relative to star 35999. Middle panel:
the same, for star 22360 (solid line, [Al/Fe]$<-0.09$) and star 31520
(dotted line, [Al/Fe]=+0.44). Lower panel: co-added spectrum of the 10 stars
with highest S/N. The areas hatched in red and light blue mark the regions used
to evaluate the flux appropriate for Al lines as discussed in the text; red is
the in-line region, light blue indicate the reference for the continuum.
}
\label{f:sintesi}
\end{figure}

In Fig.~\ref{f:sintesi} (upper panel) we compare the spectra of two stars with
very similar atmospheric parameters (effective temperature $T_{\rm eff}$,
surface gravity $\log g$, overal metal abundance [Fe/H], and microturbulent
velocity $V_t$): star 44939 (4425/1.60/-1.15/1.01) and star 35999
(4442/1.64/-1.15/0.81), with S/N ratios of 147 and 49, respectively. These are
the coolest giants in our sample. We also superimpose the synthetic 
spectrum computed with the atmospheric parameters appropriate for star 35999 and
no Al ([Al/Fe]=-9.99), to show the amount of possible contamination of Al lines
by CN features (blue thin solid line).
Another example is shown in the middle panel
of Fig.~\ref{f:sintesi}, where we compare a pair of RGB stars with similar
parameters (star 22360: 4755/2.27/-1.15/0.86 and star 31520:
4756/2.28/-1.23/1.26) and S/N ratios close to the median value for the sample
(S/N=60 and 61 for stars 22360 and 31520, respectively).

\section{Analysis}

Owing to the quality of spectra and to the weakness of Al~{\sc i} lines in
rather warm giants we decided to derive the abundances from the comparison 
of the observed flux in the region of Al lines with the flux measured on 
synthetic spectra computed using the package ROSA (Gratton 1988), with the
following procedure. 

First, to account for the contamination of CN lines\footnote{Other contaminants,
like TiO molecular lines, are not important.} present over the spectral range
where Al lines lie, we first reproduced a CN feature at $\sim 8770.5$~\AA,
adopting a fixed C abundance (0.0 dex) and varying the [N/Fe] ratio until the CN
line was reasonably well fitted. The resulting values of [N/Fe] are listed in
Table~\ref{t:tab1}. The C content adopted in this process is very likely much
higher than the actual one, which for metal-poor giants in this evolutionary
stage is expected to be about -0.6 dex. However, the meaning of [N/Fe] ratios 
is only indicative, since we lack precise abundances of C, so that these N
abundances only indicate the value that reproduce the CN feature adequately. 

Our line list, originally from B. Plez, was slightly modified by optimizing the
position and strength of CN and Al lines to fit the high resolution spectrum of
the cool and metal-rich (CN rich) giant $\mu$ Leo (see Gratton et al. 2006).

Second, we coadded the observed spectra of the 10 stars with the highest S/N in
our sample. On this spectrum we selected a region including the two Al lines and two
other regions to be used to derive a local reference continuum. These regions are 
shown as shaded red and light blue areas in Fig.~\ref{f:sintesi}. We then measured 
the average fluxes within the in-line region ($f_{\rm Al}$) and the reference 
continuum regions ($f_1$\ and $f_2$). We also estimated photometric errors in these 
regions from the S/N of the spectra and width (and then number of pixels) within 
each of these regions. Finally, we defined a line strength index for the Al
lines as $I_{\rm Al}=2~f_{\rm Al}/(f_1+f_2)$, with an error which is obtained
by a suitable combination of the photometric errors. The same procedure was
then repeated on a set of three synthetic spectra computed for each star using the
appropriate atmospheric parameters (from Carretta et al. 2011a) and abundances of
[Al/Fe]=-0.5, 0.0, 0.5. Abundances of Al for each stars were then derived by
interpolating the normalized flux in the line region among those obtained from
the synthetic spectra. An error can be attached by comparing these Al abundances
with those obtained entering a new value of the Al line strength index that is the
sum of the original value and of its error. Whenever the measured value for the
Al line strength index was smaller than twice the error, we considered that only
an upper limit to Al abundances could be obtained. In this case, we assumed that
the upper limit to Al abundances be equal to three times the error: this is a
rather robust estimate of upper limit. This procedure avoids any subjective 
judgement, as suggested by the referee, and was applied to all the observed spectra 
with a $S/N>25$: for lower $S/N$ spectra we only obtained high upper limits that
do not bear important information.

We notice that typical errors associated to the Al abundances obtained following
this approach are in the range 0.1-0.4 dex. Star-to-star errors in the adopted 
atmospheric parameters are quite small (see Carretta et al. 2011a) and less a 
source of concern. 
As discussed in Carretta et al. (2012), NLTE effects are not a source of
concern in the star-to-star analysis in NGC~6752; this is even more 
true when considering more metal-rich stars, as in the present case.

\section{Results and discussion}

The derived abundances of Al for our sample are listed in Table~\ref{t:tab1}.
We obtained [Al/Fe] ratios for 60 stars, with 49 detection and 11 upper
limits.
Other abundances (for Fe, O, Na, Mg, Si) were taken from the analysis of
Carretta et al. (2011a) and repeated for convenience in this Table. As
mentioned in the Introduction, they are available for 50 stars.

\begin{figure*}
\centering
\includegraphics[scale=0.70]{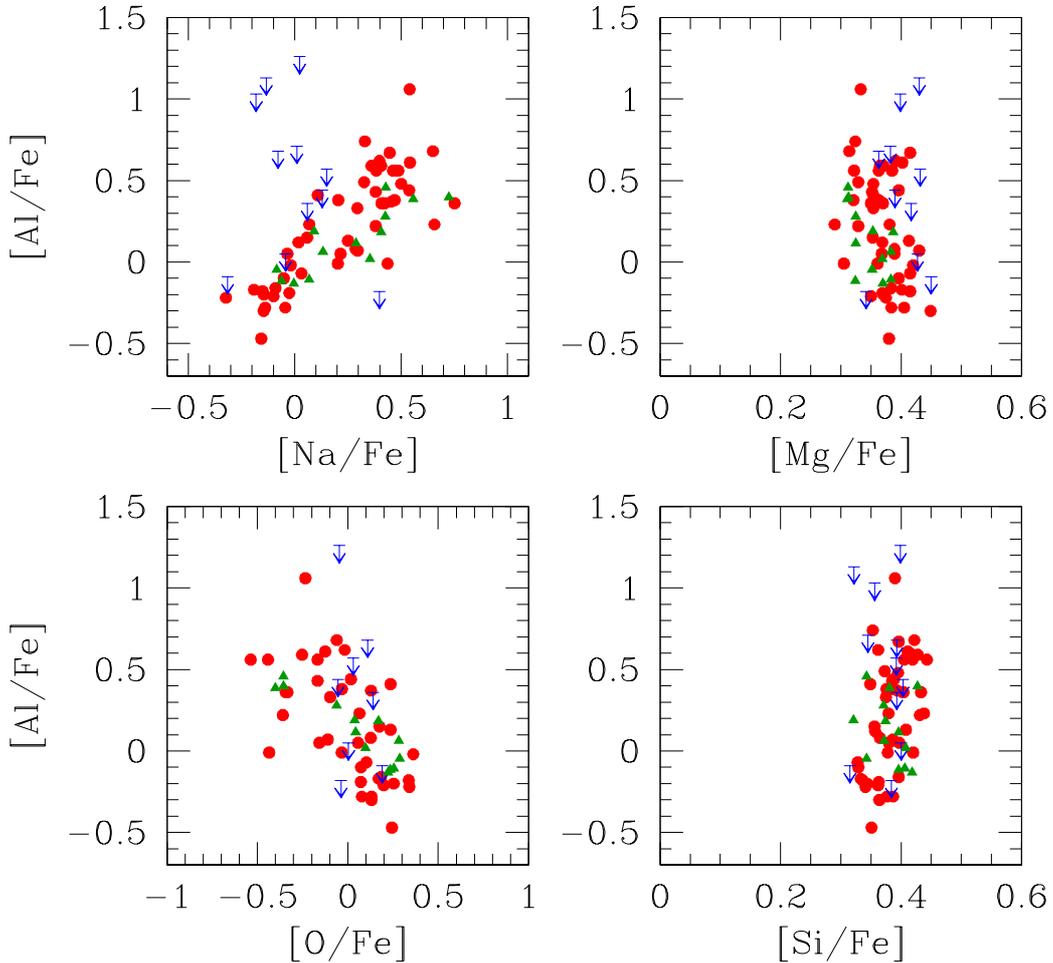}
\caption{[Al/Fe] ratios derived in this work as a function of abundances of
other proton-capture elements from Carretta et al. (2011a): [Na/Fe] (upper left
panel), [Mg/Fe] (upper right), [O/Fe] (lower left), and [Si/Fe] (lower right).
Green triangles are RGB stars with UVES spectra in NGC~1851, from
Carretta et al. (2011a). Arrows indicate upper limits in Al abundances.}
\label{f:figabu}
\end{figure*}

Al abundances do not present any significant trend as a function of
effective temperature or metallicity [Fe/H]. The expected correlations of Al
with elements enhanced by proton-capture reactions (Na, Si) and the
anticorrelations with those depleted in H-burning at high temperature (O, Mg)
are illustrated in Fig.~\ref{f:figabu}.

As a comparison - and an useful check - we add to these relations also the
abundances for 13 RGB stars with high resolution UVES spectra, with [Al/Fe]
ratios obtained from the classical Al~{\sc i} doublet at 6696-98~\AA\ in
Carretta et al. (2011a). Despite none of these 13 giants is in common with the
stars in the present sample, the nice agreement we can see in
Fig.~\ref{f:figabu} supports our abundance determination from medium-resolution
GIRAFFE spectra.

\begin{figure}
\centering
\includegraphics[scale=0.40]{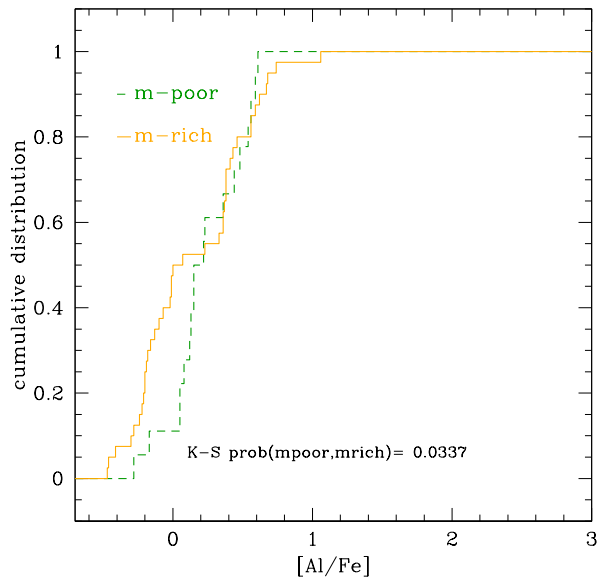}
\caption{Cumulative distribution of [Al/Fe] ratios from the present work in the
metal-poor (green dashed line) and in the metal-rich (orange solid line)
component of NGC~1851.}
\label{f:distrAl}
\end{figure}

Using a statistical cluster analysis, in Carretta et al. (2012a) we
separated within NGC~1851 two stellar components, one more
metal-rich/Ba-rich and the other more metal- and Ba-poor,
with evidence of a slightly more extreme processing among stars of the
second generation of the metal-rich component. 
In Fig.~\ref{f:distrAl} we show the cumulative distribution of Al abundance for
stars in our present sample divided into the two components: 
a statistical Kolmogorov-Smirnov test on the cumulative distributions show
evidence that the two distributions are actually different, concerning their Al
abundance, although in this case the metal-rich component seems to be more
Al-poor.

The range of [Al/Fe] ratios among giants of NGC~1851 is intermediate
between the small range observed in GCs like M~4 (e.g. Marino et al. 2008,
Carretta et al. 2009b) and the rather large range observed in massive GCs like 
NGC~2808 (Carretta et al. 2009b, Bragaglia et al. 2010).

On the other hand, Al-Si correlation observed in our GIRAFFE sample
(lower right panel in
Fig.~\ref{f:figabu}) is found to be statistically significant at a confidence
level higher than 97.5\% (Pearson's correlation coefficient $r=0.29$, 57 degree
of freedom). This correlation
demonstrates that part of the material polluting the gas
used in the formation of second generation stars was processed under
temperatures higher than about 65 MK. Indeed, this is the threshold value above
which the reaction $^{27}$Al($p$,$\gamma$)$^{28}$Si dominates on the 
$^{27}$Al($p$,$\alpha$)$^{24}$Mg reaction (see Arnould et al. 1999) and a
certain amount of $^{28}$Si is produced as a leakage from Mg-Al cycle 
(Karakas and Lattanzio 2003). The correlation between Al and Si (or the
corresponding Mg-Si anticorrelation) is now observed in a number of GCs (see
Yong et al. 2005 and Carretta et al. 2009b, 2011a) and confirmed to exist also in
NGC~1851 by the present large sample.

Finally, we found that the average abundance of Al steadily increases as the
chemical composition changes from the pattern typical of first generation stars
to that of second generation stars with increasingly modified composition.
By dividing our sample into the three components P, I, and E defined in Carretta
et al. (2009a) we find the following average values: 
[Al/Fe]$=-0.11\pm 0.06$ dex ($rms=0.25$, 17 stars) for the primordial P
component, and [Al/Fe]$=+0.36\pm 0.06$ dex ($rms=0.32$, 26 stars), and
[Al/Fe]$=+0.56\pm 0.01$ dex ($rms=0.01$, 2 stars) for the I and E 
components, respectively, of second generation stars in NGC~1851\footnote{Twelve
stars lack O abundances, therefore cannot be classified as P, I or E stars as in
Carretta et al. (2009a)}.

Looking at the relation between proton-capture and neutron-capture elements is
of particular interest in this cluster. First, Yong and Grundahl (2008) found a
correlation of Zr and La with Al in a few bright giants in NGC~1851. Carretta et
al. (2011a) confirmed this correlation using a large sample of more than 120
giants; Gratton et al. (2012b) found a close correlation between Na and Ba among
RHB stars; and finally Gratton et al. (2012c) found (i) a clearly different Sr
and Ba abundances between the faint (larger abundances) and bright subgiant 
branches (SGBs) found by Milone et al. (2008), and (ii) that a spread exists
within both sequences.

\begin{figure}
\centering
\includegraphics[scale=0.40]{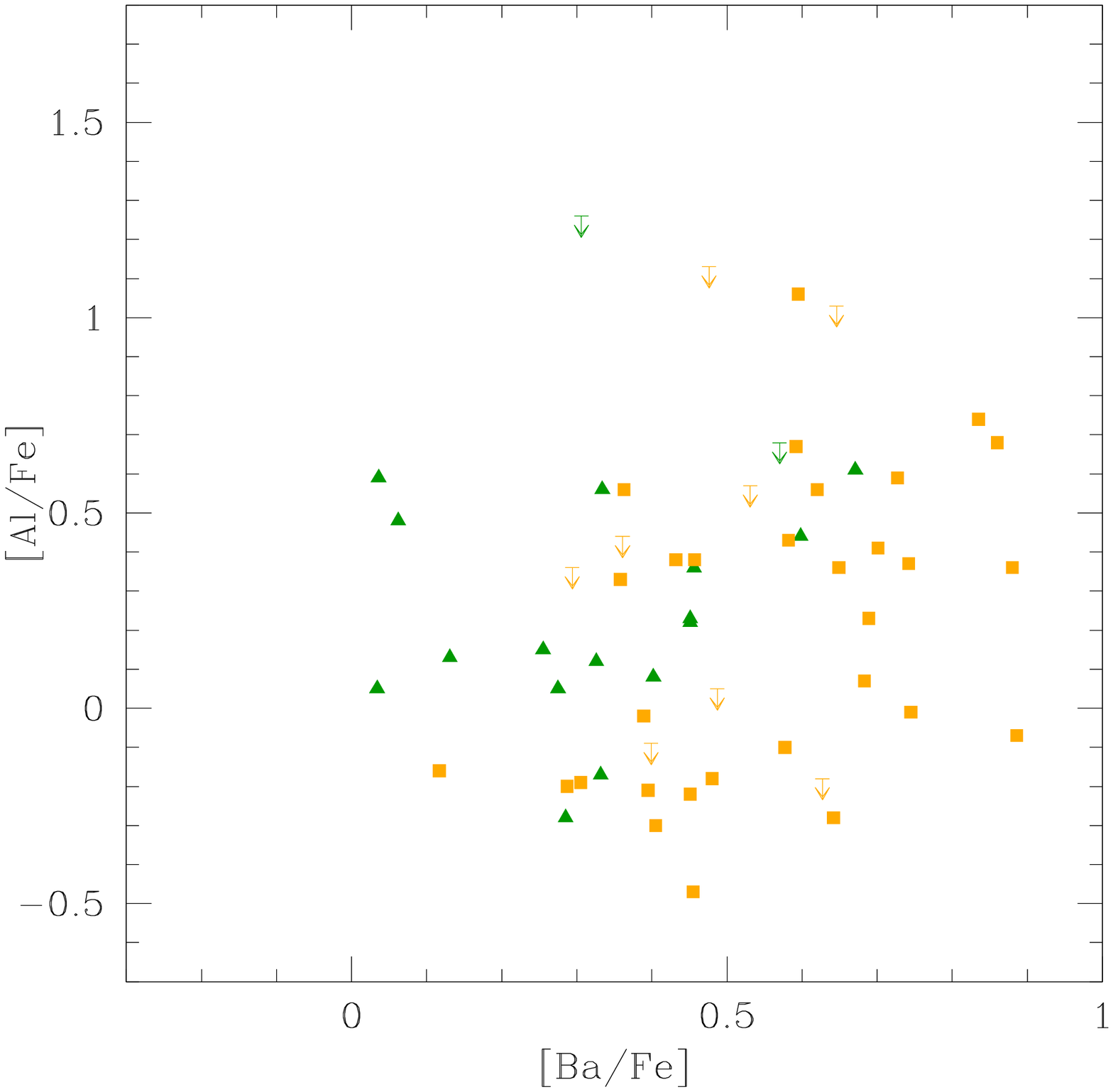}
\caption{Run of the [Al/Fe] ratio as a function of [Ba/Fe] in our sample. Green
triangles and orange squares are the metal-poor and metal-rich component,
respectively, found by Carretta et al. (2011a) among giants in NGC~1851.}
\label{f:albaclu}
\end{figure}

In Fig.~\ref{f:albaclu} we display the ratio [Al/Fe] as a function of [Ba/Fe]
from Carretta et al. (2011a) for stars in the present sample, separating the
metal-poor (green triangles) and metal-rich (orange squares) components. 

Overall, no correlation is apparent between Al and Ba; however, when
considering separately the two components, there seems to be a correlation among
the metal-rich component: the Pearson linear correlation coefficient is 0.29 for
a sample of 36 objects, which is significant to a level of confidence higher than
95\%. This finding is supported by other evidence, since both Al (this
paper) and Ba (Yong and Grundahl 2008; Carretta  et al. 2011a) are found to be
correlated with Na. 

Another confirmation comes from the relation of Al abundances with
Str\"omgren photometry. This set of filters was recently found to be very
sensitive to the  abundance of light elements such as C and N (see Carretta et
al. 2011a,b and  Sbordone et al. 2011), hence quite useful when coupled with
abundances of light  elements to explain segregations or splitting observed in
photometric sequences  in particular clusters like NGC~1851.

\begin{figure}
\centering
\includegraphics[scale=0.40]{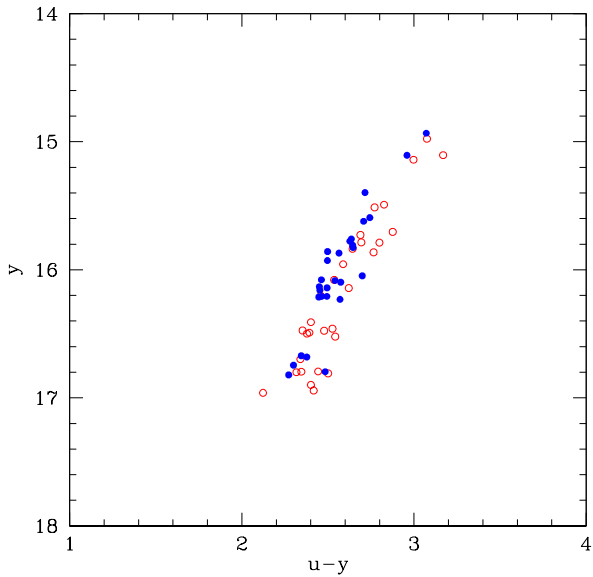}
\caption{Str\"omgren colour-magnitude diagram $y,u-y$ for giants in our sample.
Red open and blue filled symbols indicate stars with Al abundances larger and
lower, respectively, than the average ratio [Al/Fe]=0.24 dex of the sample.}
\label{f:aluy18}
\end{figure}

In Fig.~\ref{f:aluy18} we plot the $y,u-y$ CMD where the upper RGB in NGC~1851
appears well separated. On this CMD stars of our
sample are separated according to the Al abundances: red open symbols and blue
filled symbols indicate respectively giants with [Al/Fe] higher or lower than the
average value of the sample, 0.24 dex.
Almost all stars with low Al content (except for three cases) define an
extremely narrow stripe on the bluest envelope of the branch, whereas the
high-Al stars are more spread out to the red, and the well separated reddest
sequence contains almost only high Al objects, except for the three interlopers
with low Al.

In Carretta et al. (2011a,b) we also showed that this red sequence is 
preferentially populated by giants with high Ba abundances and that the
separation from the bluest sequence is mostly caused by a relevant excess in N.

Taking all these results into account, although this is not a 1 to 1
correlation, the trend of Al-rich stars to occupy preferentially the redder
sequence
suggests that, at least for the upper RGB, the Str\"omgren colours do
measure N and are not strictly correlated with the progeny of SGB stars, in agreement
with what found in Carretta et al. (2011a) using mainly O and Na abundances.
In turn, we caution that any identification of the progeny of SGB stars only
based on colour (see e.g. Han et al. 2009) must be regarded with caution,
because there is no a one-to-one correlation, hence may be misguiding.

Together with results on SGB stars by Gratton et al. (2012c) we can conclude
that stars on the faint SGB are predominantly - but non only - N-rich (hence
they occupy a redder position on the RGB, in Str\"omgren colours), while the
stars on the bright SGB are about 1/3 N-rich and 2/3 N-poor (hence they are
mostly blue in the Str\"omgren colours).

In summary, the Al abundances we obtained from spectra acquired as calibration
within the Gaia-ESO public survey add another constraint to the complex
scenario of the star formation in NGC~1851: both the putative populations hosted
in this cluster have not only a spread in Na, but also an Al spread, and the two
are quite similar in both populations.

\begin{acknowledgements}
This work was partially funded by the PRIN INAF 2009 grant CRA 1.06.12.10
(``Formation and early evolution of massive star clusters", PI. R. Gratton), 
and by the PRIN INAF 2011 grant ``Multiple populations in globular clusters:
their role in the Galaxy assembly" (PI E. Carretta). We thank 
\v{S}ar\={u}nas Mikolaitis for sharing with us the line list for CN provided 
by B. Plez. We also thank the referee for her/his constructive suggestions and
for asking for a less subjective method,
since this resulted in better constrained and accurate abundances.
This research has made use of the SIMBAD database (in particular  Vizier),
operated at CDS, Strasbourg, France and of NASA's Astrophysical Data System.
\end{acknowledgements}

\clearpage
\begin{table*}
\centering
\caption[]{Relevant information and derived abundances for our sample of RGB stars in NGC~1851.}
\begin{tabular}{rcrrrrrrrrrrrr}
\hline
obj$^1$ & star$^2$&S/N$^3$& [Al/Fe]$^3$&err&lim$^3$&[N/Fe]$^3$& $B^2$ & $V^2$  &[Fe/H]$^2$&[O/Fe]$^2$& [Na/Fe]$^2$&[Mg/Fe]$^2$&[Si/Fe]$^2$\\
\hline
 541 & 13618 &  35 & -0.01&  0.46 & 1  &   -0.05& 16.392& 15.482 & -1.120& -0.435&  0.437&  0.305&  0.420  \\ 
1461 & 14827 &  32 &  0.74&  0.36 & 1  &        & 17.819& 16.994 & -1.164&       &  0.330&  0.324&  0.353  \\ 
 974 & 16120 &  98 &  0.41&  0.09 & 1  &    0.10& 16.859& 16.012 & -1.103&  0.236&  0.109&  0.354&  0.349  \\ 
1218 & 20189 &  34 &  0.44&       & 0  &    0.70& 17.339& 16.521 & -1.155& -0.055&  0.129&  0.390&  0.403  \\ 
1186 & 20426 &  38 &  0.36&       & 0  &    0.30& 17.281& 16.429 & -1.106&  0.139&  0.060&  0.417&  0.393  \\ 
1044 & 20653 &  35 &  0.08&  0.28 & 1  &    0.55& 17.002& 16.120 & -1.184&  0.127&  0.287&  0.389&  0.365  \\ 
 733 & 20922 &  41 &  0.05&       & 0  &    0.10& 16.705& 15.784 & -1.142&  0.004& -0.043&  0.427&  0.400  \\ 
 766 & 21453 &  60 &  0.05&  0.15 & 1  &    0.25& 16.746& 15.835 & -1.156& -0.157&  0.216&  0.368&  0.380  \\ 
1211 & 21830 &  45 &  0.38&  0.22 & 1  &    0.50& 17.344& 16.515 & -1.163& -0.033&  0.470&  0.363&  0.388  \\ 
1103 & 22360 &  60 & -0.09&       & 0  &    0.10& 17.123& 16.223 & -1.145&  0.191& -0.315&  0.450&  0.315  \\ 
 788 & 22588 &  80 & -0.28&  0.23 & 1  &    0.10& 16.762& 15.878 & -1.200&  0.078& -0.139&  0.405&  0.377  \\ 
 891 & 22813 & 120 & -0.30&  0.13 & 1  &    0.20& 16.820& 15.945 & -1.101&  0.131& -0.146&  0.449&  0.364  \\ 
 719 & 23647 &  57 &  0.22&  0.16 & 1  &    0.00& 16.693& 15.806 & -1.193& -0.359&  0.381&  0.329&  0.431  \\ 
 810 & 23765 &  61 &  0.36&  0.14 & 1  &    0.05& 16.765& 15.814 & -1.145& -0.344&  0.409&  0.370&  0.403  \\ 
1016 & 25037 &  63 & -0.28&  0.29 & 1  &    0.30& 16.975& 16.106 & -1.151&  0.131& -0.044&  0.384&  0.387  \\ 
1202 & 25799 &  38 &  0.68&       & 0  &        & 17.344& 16.493 & -1.214&  0.109& -0.078&  0.363&  0.393  \\ 
1190 & 26532 &  41 &  0.68&  0.24 & 1  &        & 17.293& 16.472 & -1.180& -0.061&  0.650&  0.314&  0.422  \\ 
1045 & 26552 &  57 & -0.18&       & 0  &    0.10& 17.019& 16.090 & -1.114& -0.038&  0.399&  0.342&  0.384  \\ 
1291 & 26880 &  57 & -0.18&  0.33 & 1  &    0.30& 17.537& 16.699 & -1.161&  0.337& -0.151&  0.415&  0.336  \\ 
1274 & 27491 &  43 & -0.01&  0.48 & 1  &    0.40& 17.499& 16.685 & -1.177& -0.034&  0.203&  0.361&  0.378  \\ 
1070 & 28116 &  32 &  0.71&       & 0  &    0.40& 17.040& 16.200 & -1.189&       &  0.010&  0.382&  0.345  \\ 
 418 & 29203 &  57 &  0.56&  0.14 & 1  &    0.10& 16.162& 15.115 & -1.157& -0.442&  0.487&  0.363&  0.443  \\ 
1232 & 29470 &  41 &  0.59&  0.25 & 1  &    0.40& 17.413& 16.548 & -1.144&       &  0.361&  0.371&  0.428  \\ 
1024 & 30286 &  28 &  0.57&       & 0  &    0.20& 16.985& 16.115 & -1.072&  0.031&  0.151&  0.432&  0.392  \\ 
 802 & 31284 &  75 &  0.05&  0.12 & 1  &    0.00& 16.767& 15.893 & -1.179&  0.057& -0.034&  0.389&  0.397  \\ 
1307 & 31399 &  48 & -0.02&  0.38 & 1  &        & 17.569& 16.761 & -1.158&  0.362& -0.018&  0.420&         \\ 
1410 & 31463 &  30 &  1.13&       & 0  &        & 17.774& 17.031 & -1.163&       & -0.134&  0.430&  0.322  \\ 
1046 & 31520 &  61 &  0.44&  0.16 & 1  &    0.00& 16.998& 16.158 & -1.230&  0.017&  0.539&  0.396&  0.388  \\ 
 553 & 32112 &  43 &  0.43&  0.21 & 1  &    0.30& 16.460& 15.517 & -1.077& -0.167&  0.381&  0.352&  0.385  \\ 
1012 & 32256 &  44 & -0.10&  0.38 & 1  &        & 16.947& 16.109 & -1.168&  0.073& -0.051&  0.396&  0.329  \\ 
 875 & 35750 &  66 &  0.36&  0.13 & 1  &    0.20& 16.813& 15.874 & -1.185& -0.334&  0.423&  0.356&  0.404  \\ 
 347 & 35999 &  49 &  1.06&  0.18 & 1  &    0.00& 15.962& 14.971 & -1.147& -0.235&  0.541&  0.333&  0.390  \\ 
1395 & 36292 &  27 &  0.37&  0.33 & 1  &        & 17.724& 16.917 & -1.101&  0.128&  0.454&  0.365&  0.394  \\ 
 757 & 38484 &  29 &  0.61&  0.30 & 1  &        & 16.727& 15.837 & -1.198& -0.125&  0.542&  0.402&  0.411  \\ 
1324 & 38818 &  40 &  0.23&  0.24 & 1  &    0.00& 17.618& 16.804 & -1.144&       &  0.657&  0.290&  0.438  \\ 
 801 & 39364 &  54 &  0.33&  0.16 & 1  &    0.20& 16.765& 15.877 & -1.128& -0.098&  0.295&  0.354&  0.375  \\ 
 539 & 40300 & 124 &  0.49&  0.07 & 1  &   -0.10& 16.414& 15.533 & -1.192&       &  0.327&  0.329&  0.373  \\ 
 409 & 40615 &  72 &  0.59&  0.12 & 1  &   -0.20& 16.114& 15.156 & -1.231& -0.253&  0.407&  0.365&  0.419  \\ 
 692 & 41113 &  68 &  0.56&  0.14 & 1  &    0.20& 16.642& 15.769 & -1.219& -0.537&  0.463&  0.322&  0.419  \\ 
1323 & 41855 &  35 &  0.48&  0.30 & 1  &        & 17.610& 16.847 & -1.202&       &  0.501&  0.354&  0.395  \\ 
1339 & 43528 &  41 & -0.22&  0.45 & 1  &    0.40& 17.624& 16.854 & -1.067&  0.340& -0.323&  0.375&  0.341  \\ 
1351 & 44224 &  38 &  0.36&  0.27 & 1  &        & 17.660& 16.839 & -1.205&       &  0.752&  0.350&  0.433  \\ 
 616 & 44414 &  56 &  0.23&  0.18 & 1  &    0.10& 16.549& 15.653 & -1.187&  0.064&  0.069&  0.381&  0.379  \\ 
 346 & 44939 & 147 & -0.19&  0.10 & 1  &   -0.20& 15.952& 14.956 & -1.151&  0.072& -0.025&  0.369&  0.363  \\ 
1119 & 45006 &  63 & -0.07&  0.28 & 1  &    0.00& 17.096& 16.245 & -1.115&  0.103&  0.032&  0.415&  0.328  \\ 
1080 & 45090 &  59 &  0.15&  0.16 & 1  &        & 17.047& 16.228 & -1.185&  0.175&  0.059&  0.353&  0.356  \\ 
1050 & 45413 &  88 & -0.47&  0.21 & 1  &    0.30& 16.981& 16.155 & -1.125&  0.244& -0.157&  0.380&  0.351  \\ 
1093 & 46228 &  73 & -0.21&  0.21 & 1  &        & 17.075& 16.238 & -1.097&  0.198& -0.099&  0.350&  0.362  \\ 
1203 & 46657 &  30 &  1.26&       & 0  &        & 17.316& 16.521 & -1.246& -0.047&  0.023&       &  0.399  \\ 
1074 & 46958 &  93 & -0.20&  0.20 & 1  &    0.25& 17.048& 16.204 & -1.142&  0.254& -0.145&  0.371&  0.344  \\ 
 507 & 47385 &  66 & -0.17&  0.27 & 1  &    0.00& 16.341& 15.437 & -1.203&  0.171& -0.191&  0.401&  0.333  \\ 
 746 & 47795 &  38 &  0.56&  0.23 & 1  &    0.40& 16.723& 15.753 & -1.127& -0.167&  0.384&  0.385&  0.405  \\ 
 407 & 48085 & 119 & -0.16&  0.11 & 1  &   -0.20& 16.115& 15.144 & -1.111&  0.183& -0.090&  0.384&  0.396  \\ 
1350 & 48277 &  26 &  1.03&       & 0  &        & 17.659& 16.863 & -1.068&       & -0.181&  0.399&  0.356  \\ 
1007 & 48388 &  58 &  0.12&  0.16 & 1  &    0.00& 16.937& 16.136 & -1.163&       &  0.018&  0.369&  0.357  \\ 
1144 & 49965 &  45 &  0.62&  0.20 & 1  &    0.30& 17.156& 16.242 & -1.095& -0.017&  0.399&  0.391&  0.362  \\ 
1280 & 50876 &  37 &  0.38&  0.23 & 1  &    0.30& 17.512& 16.680 & -1.097&       &  0.205&  0.321&  0.376  \\ 
 588 & 50973 &  64 &  0.07&  0.14 & 1  &   -0.20& 16.530& 15.572 & -1.084& -0.111&  0.296&  0.430&  0.386  \\ 
 738 & 51311 &  62 &  0.13&  0.16 & 1  &    0.30& 16.706& 15.758 & -1.256&  0.236&  0.251&  0.413&  0.408  \\ 
1319 & 52579 &  33 &  0.67&  0.28 & 1  &        & 17.587& 16.801 & -1.151&       &  0.447&  0.415&  0.396  \\ 
\hline
\end{tabular}
\label{t:tab1}
\begin{list}{}{}
\item[1-] object identification, from FLAMES mask in the ESO archive
\item[2-] star identification, magnitudes and abundance ratios taken from Carretta et al. (2011)
\item[3-] S/N ratios and abundances derived in the present work. lim is 0 for upper limits in the Al abundance.
\end{list}

\end{table*}

\end{document}